## The CONQUEST code: large scale and linear scaling DFT

D. R. Bowler[1,2], T. Miyazaki[2], A. Nakata[2] and L. Truflandier[3]
[1]London Centre for Nanotechnology, London, UK
[2]WPI-MANA, National Institute for Materials Science, Tsukuba, Japan
[3]Institut des Sciences Moleculaires, Université Bordeaux, Bordeaux, France

**Background and Current Status**

CONQUEST[1] is a DFT code which was designed from the beginning to enable extremely large-scale calculations on massively parallel platforms, implementing both exact and linear scaling solvers for the ground state. It uses local basis sets (both pseudo-atomic orbitals, PAOs,[2] and systematically convergent B-splines[3]) and sparse matrix storage and operations to ensure locality in all aspects of the calculation.

Using exact diagonalisation approaches and a full PAO basis set, systems of up to 1,000 atoms can be modelled with relatively modest resources (200-500 cores), while use of multi-site support functions (MSSF)[4] enable calculations of up to 10,000 atoms with similar resources. With linear scaling, the code demonstrates essentially perfect weak scaling (fixed atoms per process), and has been applied to over 1,000,000 atoms, scaling to nearly 200,000 cores[5]; it has been run on both the K computer and Fugaku, among other computers.

CONQUEST calculates the total energy, forces and stresses exactly, and allows structural optimisation of both ions and simulation cell. Molecular dynamics calculations within the NVE, NVT and NPT ensembles are possible with both exact diagonalisation and linear scaling[6]. The code interfaces with LibXC to implement LDA and GGA functionals, with metaGGA and hybrid functionals under development. Dispersion interactions can be included using semi-empirical methods (DFT-D2/3, TS) and vdW-DF. The polarisation can be calculated using Resta's approach.

We have recently applied CONQUEST to calculations in complex ferroelectric systems with up to 5,000 atoms[7,8,9], investigating problems that require large simulation cells and electronic structure methods. In Fig. 1a) and b), the local polarisation textures of PbTiO3 thin films on SrTiO3[7] are shown for two thicknesses of film: 9 layers (top) and 3 layers (bottom). The formation of polar vortices is clear in the thick film, while the thinner film cannot support these, instead showing a polar wave with chiral bubbles forming at the surface; we have extended these studies to investigate the interaction of domain walls with surface trenches[8]. In Fig 1c), we plot the partial charge density from the conduction band minimum (CBM) in YGaO3, superimposed on a map of the tilt angle of the GaO5 bipyramids relative to the (001) direction, which shows domain walls. Domain wall meeting points (dark blue) are topologically protected, and show a concentration of the CBM, reflecting a reduced band gap[9].

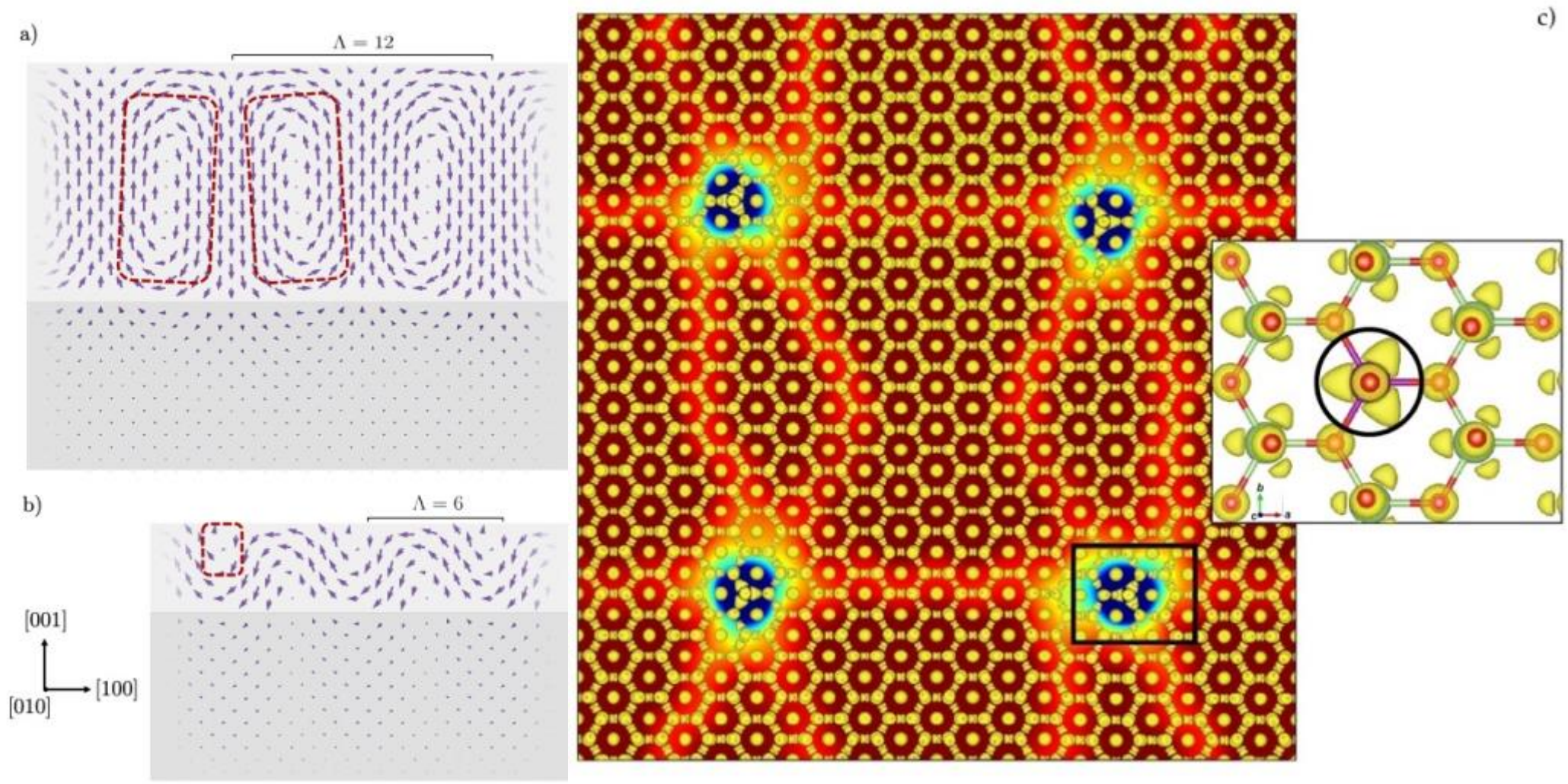

**Figure 1.** (a, b) Polarisation textures in $PbTiO_3$ films on $SrTiO_3$[7]. (c) Charge density from conduction band minimum in $YGaO_3$, above a map of tilt angle in $GaO_5$ bipyramids. Reproduced from Phys. Rev. B 102, 144103 (2020) with permission.

## Development Priorities

CONQUEST already allows simulation of significantly larger systems than standard DFT codes, and we want to improve the efficiency and scaling of the code to enable even larger systems to be modelled. We also want to reduce the total computational time required per MD step, to enable longer timescales to be addressed. At present, exact diagonalisation simulations use the ScaLAPACK solver, but an interface to ELPA has been developed and is being deployed. Alongside these efficient approaches to exact diagonalisation, we are prioritising lower scaling solvers which allow the calculation of selected eigenstates.

Perhaps the best known of these in the electronic structure community is PEXSI, which is available through ELSI, though it would require some work to interface this to the CONQUEST matrix storage (which was designed specifically for high efficiency, highly parallel linear scaling calculations). We already have a post-hoc interface to an implementation of the Sakurai-Sugiura method, which scales as O(mN^{1-2}) when finding m eigenstates for N atoms, and shows extremely good parallel scaling, and we will investigate the possibility of incorporating this approach into CONQUEST as an alternative solver.

For the linear scaling solver, we will improve both the accuracy and the robustness. The key limitation on accuracy at present is the suitability of basis sets: the blip functions are accurate, and systematically improvable, but can be slow to converge, while PAOs are limited to relatively small sizes (typically single zeta plus polarisation, or SZP). The improvement of blip optimisation will concentrate on two aspects: first, the search methods used for the blips themselves; second, the integration with the linear scaling optimisation of the density matrix. For PAO basis sets, we will continue to develop an on-site equivalent of the MSSF approach, and develop an extension to full MSSF. We will also investigate alternative methods for linear scaling inversion of the overlap matrix, which is key to efficient linear scaling solution, and is sensitive to the basis set.

Efficient methods for metallic or small gap systems are also extremely important. The linear scaling solver in CONQUEST is not suitable for these systems, so we will investigate alternative approaches, including iterative methods. The question of large-scale, efficient solvers for these systems is one that is still of paramount importance to the large-scale DFT community as a whole.

We are in the process of including metaGGA functionals into CONQUEST, and will continue this alongside a robust, efficient implementation of exact exchange (for which we already have preliminary results). We will also enable solutions with spin-orbit coupling and the full Dirac equation. We have reported the successful linear scaling implementation of real-time TDDFT, and will extend this to the exact solvers, including the Casida linear-response approach for the exact solvers. We will also implement density-matrix perturbation theory (DMPT), as required by DFPT, for both linear scaling and exact solvers, to enable response function calculations to be performed.

**Meeting the Exascale Challenges**

The key challenge associated with the transition to exascale computing for CONQUEST is to update and adapt the code to new hardware architectures, in particular, efficient use of CPUs with many cores, and GPUs and other co-processors, while maintaining excellent parallel scaling. The code will need to become more heterogeneous, with on-node calculations distributed between CPU cores and GPUs, and a few MPI processes dedicated to inter-node communication; careful interleaving of different parts of the calculation to different hardware on a node will be key to maximising performance.

When working in exact diagonalisation mode, CONQUEST currently relies on ScaLAPACK, with an interface to ELPA in deployment. We will monitor developments in the area of exascale solvers (including projects such as MAGMA), as well as implementing other solvers which scale well in parallel, and which can make efficient use of local resources.

CONQUEST has, so far, shown no issues with parallel scaling when used in the weak scaling mode, even as far as 200,000 processes, as illustrated in Fig. 2. We are aware of areas in the code which might start to pose problems at larger process counts; the key area is the use of FFTs, though these are only used for the Hartree potential, calculation of gradients of the charge density, and for vdW-DF functionals. There are well-established alternatives for the first two of these operations which we will implement as necessary. The other area which may offer issues is the storage of atomic positions, velocities and forces, which are currently held globally on all processes, but can be made local (to the process responsible) if needed.

The overlap of communication and calculation is an approach that is inherently possible in much of the CONQUEST code[5], though has not been extensively implemented. This overlap fits well with multi-threading and heterogeneous hardware, and will be important to good on-node exascale performance. We will expand the areas of the code that are multithreaded (using OpenMP) as well as developing GPU implementations for numerically intensive parts of the calculation. At present, we have a preliminary GPU implementation for matrix multiplication, and a full GPU implementation of the O(N) solver would offer significant advantages. We will also test the off-loading of other parts, including force calculations and mapping from density matrix to charge density on the grid, along with the EXX implementation.

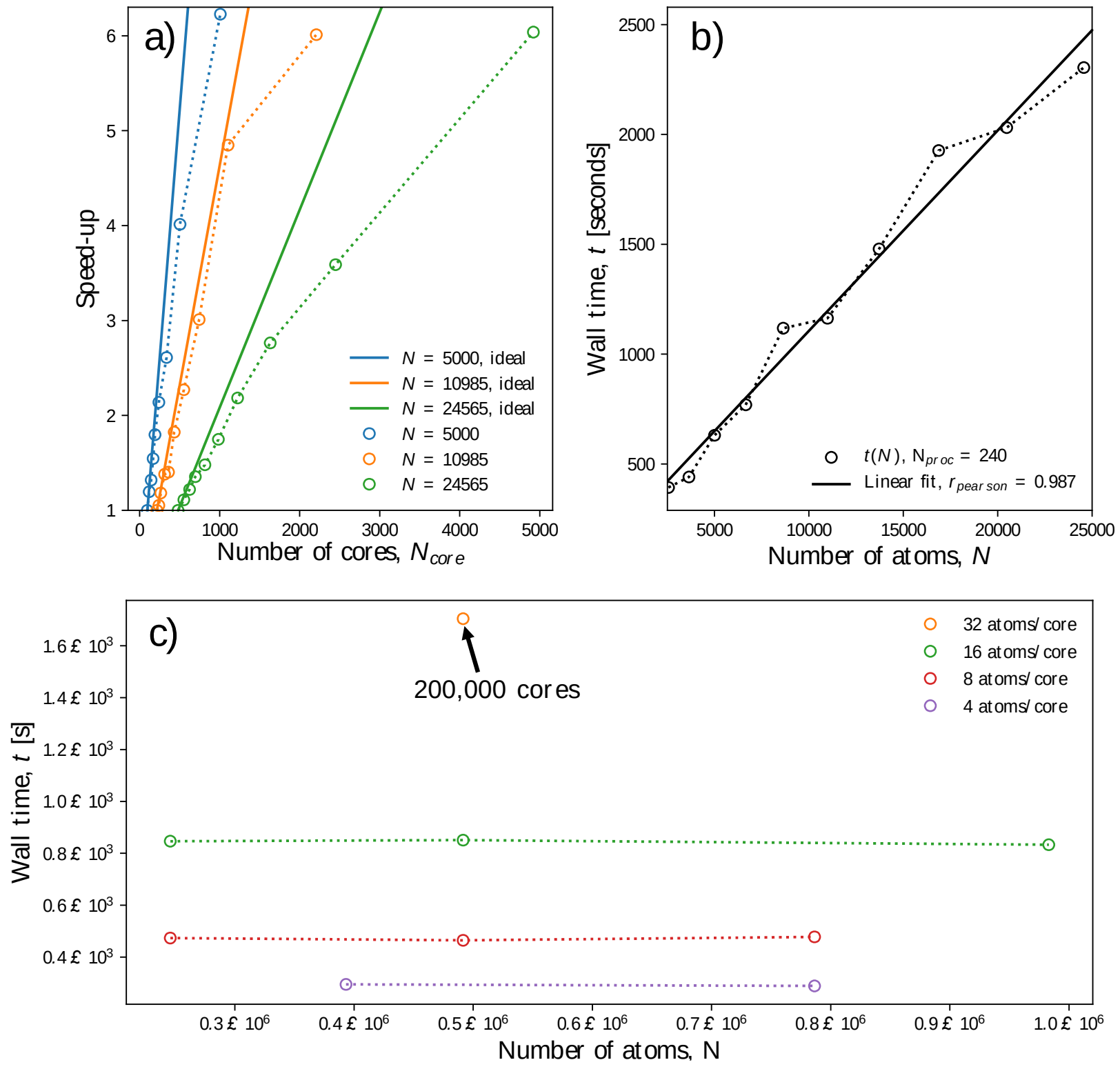

Figure 2. (a) Strong scaling for bulk PbTiO3 on the UK ARCHER computer; (b) linear scaling of computational time for the same system; (c) weak scaling on the K computer for systems up to 1,000,000 atoms of silicon. Reproduced from J. Chem. Phys. 152 164112 (2020) with permission.

## Concluding Remarks

The search for new functional materials with complex structures is a key target for exa-scale computing, and the efficient use of massively parallel DFT calculations is an important part of this search. The CONQUEST code enables large-scale exact DFT simulations with relatively modest hardware resources, paving the way for large numbers of calculations on large simulations cells on exa-scale hardware. At the same time, with linear scaling, it is capable of modelling systems with many millions of atoms, hence applying DFT to systems of experimentally relevant size in many different disciplines. With improvements to solution time, long timescales and efficient structural relaxation will become widely available for these very large systems.

## Acknowledgements

*The authors would like to acknowledge the many contributions of Professor Mike Gillan, who founded and led the CONQUEST project; we are also grateful to the other CONQUEST developers and users for all their input. The study relied on computational support from the UK Materials and Molecular Modelling Hub, which is partially funded by EPSRC (EP/P020194), for which access was obtained via the UKCP consortium and funded by EPSRC Grant Ref. No. EP/P022561/1. This work also used the ARCHER UK National Supercomputing Service funded by the UKCP consortium EPSRC Grant Ref. No. EP/P022561/1. This work is also partially supported by the projects JSPS Grant-in-Aid for Transformative Research Areas (A) "Hyper-Ordered Structures Science" (Grant Nos. 20H05883 and 20H05878) and JSPS Grant-in-Aid for Scientific Research (Grant Nos. 18H01143, 21H01008). Calculations were performed also on the Numerical Materials Simulator at NIMS in Japan.*